\documentclass[%
 reprint,
 amsmath,amssymb,
 aps,
]{revtex4-1}

\usepackage{graphicx}
\usepackage{epstopdf}
\usepackage{dcolumn}
\usepackage{bm}

\usepackage{color}

\begin{document}

\title{Field-free superconducting diode in a magnetically nanostructured superconductor}

\author{Ji Jiang$^{1,2}$, M. V. Milo\v{s}evi\'{c}$^{2}$, Yong-Lei Wang$^{3}$, Zhi-Li Xiao$^{4}$, F. M. Peeters$^{2}$,}
\author{Qing-Hu Chen$^{1,5,}$}
\email{qhchen@zju.edu.cn}

\address{
$^{1}$Zhejiang Province Key Laboratory of Quantum Technology and Device, School of Physics, Zhejiang University, Hangzhou 310027, China \\
$^{2}$Department of Physics, University of Antwerp, 2020 Antwerp, Belgium \\
$^{3}$Research Institute of Superconductor Electronics, School of Electronic Science and Engineering, Nanjing University, Nanjing 210093, China \\
$^{4}$Department of Physics, Northern Illinois University, DeKalb, Illinois 60115, USA \\
$^{5}$Collaborative Innovation Center of Advanced Microstructures, Nanjing University, Nanjing 210093, China
}

\date{\today }

\begin{abstract}
A strong superconducting diode effect (SDE) is  revealed in a thin superconducting film periodically nanostructured with magnetic dots. The SDE is caused by the current-activated dissipation mitigated by vortex-antivortex pairs (VAPs), which periodically nucleate under the dots, move and annihilate in the superconductor - eventually driving the system to the high-resistive state. Inversing the polarity of the applied current destimulates the nucleation of VAPs, the system remains superconducting up to far larger currents, leading to the pronounced diodic response. Our dissipative Ginzburg-Landau simulations detail the involved processes, and provide reliable geometric and parametric ranges for the experimental realization of such a non-volatile superconducting diode, which operates in absence of any applied magnetic field while being fluxonic by design.
\end{abstract}

\maketitle

\section{Introduction}

The modern electronics industry relies heavily on the nonreciprocal charge-transport property of semiconductor devices. One well-known example is the celebrated diode effect characterized by polarity-dependent current-voltage relation. It has been widely used in many important devices such as rectifiers~\cite{RIDEOUT1978261}, light-emitting diodes~\cite{1991led}, and solar cells~\cite{solarcells}. A hybrid structure, namely p-n junction, is usually designed to fabricate a semiconductor diode. The diode effect has also been observed in noncentrosymmetric semiconductor materials under external magnetic fields ~\cite{PhysRevLett.87.236602,PhysRevLett.94.016601}. However, in all semiconductor devices, the energy loss is inevitable even in the low resistance due to the finite barrier for the charge carriers.

The nonreciprocal charge transport in superconductors has also attracted a lot of attention~\cite{sciadv.1602390,natcommun14465,yasuda2019nonreciprocal,PhysRevB.98.054510,%
lustikova2018vortex,PhysRevLett.99.067004}. Of particular interest, superconductors can carry currents without dissipation, thus a nonreciprocal critical current due to the superconductor-metal transition could bring about the possibility of the so-called superconducting diode effect (SDE), which exhibits zero resistance in only one current direction. The SDE would not only pave the way to dissipationless electronic devices but also grant access to the fundamental mechanism of various nonreciprocal properties.

Recently, the SDE was demonstrated in an artificial superlattice $\rm [Nb/V/Ta]_{n}$ without a center of inversion under a magnetic field ~\cite{ando2020observation}. It has boosted a plethora of the studies immediately. In the past year, it has been reported that the SDE can be observed in several superconducting materials with different structures such as synthetic Josephson junctions ~\cite{baumgartner2021josephson,wu2021realization}, metal-ferromagnetic insulator-superconductor tunnel junction ~\cite{strambini2021rectification}, superconducting films patterned with a conformal array of nanoscale holes ~\cite{lyu2021superconducting}, and few-layer superconducting nanowires ~\cite{bauriedl2021supercurrent}. In these experiments, the spatial inversion symmetry is broken in different ways, e.g., stacking different superconducting layers~\cite{ando2020observation}, the spin-orbit interaction in Josephson junctions~\cite{baumgartner2021josephson}, asymmetric junction structures  ~\cite{wu2021realization,strambini2021rectification}, noncentrosymmetric pinning arrays~\cite{lyu2021superconducting}, and odd layer number ~\cite{bauriedl2021supercurrent}. {  These experimental studies have simulated many theoretical works related to the demonstration of SDE~\cite{yuan2021supercurrent,daido2021intrinsic,he2021phenomenological,zinkl2021symmetry,PhysRevB.103.144520,PhysRevB.103.245302,PhysRevB.105.104508,davydova2022universal,zhang2021general,zhai2022prediction,scammell2022theory,PhysRevLett.128.177001,tanaka2022theory}. Some of them, which are performed analytically in the framework of the Ginzburg-Landau theory under the mean-field level, focus on strong spin-orbital coupling~\cite{yuan2021supercurrent,daido2021intrinsic,he2021phenomenological} while Ref.~\cite{zinkl2021symmetry} analyzes spontaneous edge currents and the energy bias for the formation of Josephson vortices the chiral p-wave superconductors.}

Among most superconducting structures to realize the SDE, an external magnetic field is { generally} required to break the time-reversal symmetry, e.g., an in- or out-of-plane magnetic field is applied in Refs. ~\cite{ando2020observation,lyu2021superconducting,baumgartner2021josephson,bauriedl2021supercurrent}. {  Interestingly, a field-free superconducting diode is reported in an inversion-symmetry-breaking heterostructure~\cite{wu2021realization} and in EuS tunnel junctions~\cite{strambini2021rectification}. However, the possible SDE in conventional superconductors without external fields or at zero net field is rarely visited. If the SDE could be realized in a field-free conventional superconductor,} it would be a more convenient and economical approach for the development of superconducting devices. Additionally, although the nonreciprocal transport in superconductors without an inversion symmetry is fabulous, the possibility of realizing  the SDE in homogeneous conventional superconductors is much less explored. SDE that does not depend on specific materials is thrilling and can be critical to industrial production.

In this work, we propose a field-free SDE system of a two-dimensional type-II superconducting film with magnetic dots (MDs) \cite{PhysRevLett.90.197006,PhysRevLett.93.267006,VELEZ20082547,PhysRevLett.95.147004,MILOSEVIC2006208,PhysRevB.76.184516,Milo_evi__2011} on top of it, where the noncentrosymmetry superconductivity is absent. Each MD is parallel to the superconducting film, thus exhibiting a zero flux. In the absence of an external magnetic field, vortices can be topologically excited only by external currents or thermal fluctuations, namely vortex loops in three dimensions ~\cite{blatter1994RevModPhys} or vortex-antivortex pairs (VAPs) in two dimensions ~\cite{KT1973,PhysRevLett.40.783,PhysRevLett.87.067001}. In particular, in the two-dimensional case, which is highly relevant to the superconducting films, vortices move transversely to an applied current, yielding a finite voltage. Note that the correlated drift of a vortex-antivortex pair does not contribute to dissipative voltage, only the opposite motions of vortices and antivortices can produce dissipations ~\cite{KT1973,PhysRevLett.40.783,PhysRevLett.87.067001}. Since the dynamics of the same number of vortices and antivortices under zero net fields should be quite different from the vortex dynamics without antivortices, the present mechanism of SDE is distinct from the previous ones under an external magnetic field.

The paper is organized as follows. In Sec. II, we give a brief introduction to the system under investigation and the numerical method based on the time-dependent Ginzburg-Landau (TDGL) equations and the heat-transfer equation. In Secs. III and IV, the main results are presented, and the underlying mechanisms for the SDE and the reversed SDE are analyzed and revealed from the perspective of vortex dynamics. Finally, conclusions are drawn in the last section. Visualizations of the vortex dynamics and temperature evolution are provided within the Supplemental Materials \cite{supplemental}.

\section{Model and Numerical method}

\begin{figure}[htbp]
\centerline{
\includegraphics[scale=0.41]{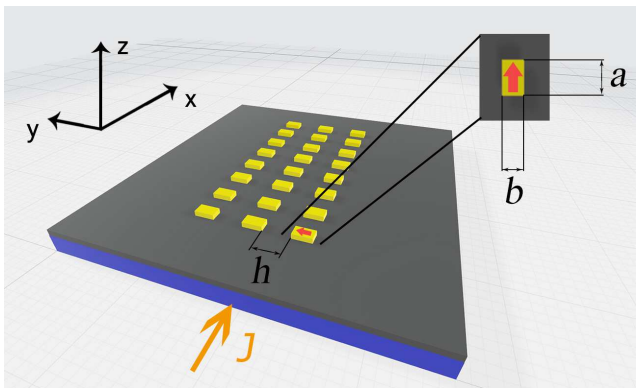}}
\caption{Schematic view of the superconducting film (blue) with rows of MDs (yellow) on top of it. A thin layer of insulating dioxide (gray), whose thickness is $0.5\protect\xi (0)$, is coated on top of the superconductor. The size of a MD is  $a=8\xi(0)$ in the length and $b=5\xi(0)$ in the width. All magnetic dots have the same thickness $1\xi(0)$. The magnetization direction of a MD is indicated by the red arrow, which is perpendicular to the applied current density $J$. In the simulations, the external current is applied along $x$ direction where the periodic boundary condition is also employed. The superconductor-vacuum boundary condition is used in the $y$ direction.}
\label{fig_sys}
\end{figure}

A schematic view of the superconducting film with MDs on top of it is shown in Fig. \ref{fig_sys}. To avoid unwanted proximity effects, they are separated by an insulating layer. The famous phenomenological Ginzburg-Landau theory can describe the nonuniform superconductors well ~\cite{tinkham2004introduction,Schmid1966ATD,WOS:A1968C188700031} and the TDGL theory can be employed to simulate the dynamics of various hybrid superconducting systems \cite{
PhysRevLett.94.227001,PhysRevB.80.214509,PhysRevB.81.054503,PhysRevB.80.054503,PhysRevB.86.224504}. It has been extensively reported in the literature that the TDGL simulation results are in excellent agreement with the experimental observations both in homogeneous magnetic fields~\cite{PhysRevB.95.075303,PhysRevLett.109.057004,PhysRevLett.111.067001} and in inhomogeneous fields~\cite{PhysRevLett.126.117205}.

TDGL equations in the dimensionless form are given by Refs. \cite{WINIECKI2002127,PhysRevB.86.224504,PhysRevLett.109.057004}:
\begin{equation}
u(\frac{\partial \psi }{\partial t} + i \phi \psi)=(\nabla -i\mathbf{A}%
)^{2}\psi +(1 - T -|\psi|^{2}) \psi +\chi (\mathbf{r},t),  \label{tdgl1}
\end{equation}%
\begin{equation}
\sigma \frac{\partial \mathbf{A}}{\partial t} + \nabla \phi=Im(\psi ^{\ast
}(\nabla -i \mathbf{A})\psi )-\kappa ^{2}\nabla \times \nabla \times \mathbf{%
A},  \label{tdgl2}
\end{equation}%
where $\psi $, $\mathbf{A}$, $\phi$ and $T$ are the superconducting order parameter, the vector potential, the electric potential and local temperature, respectively. $u$ is a phenomenological coefficient and usually accepted as $u=1$ ~\cite{PhysRevB.84.094527,PhysRevB.86.224504}. Distance is measured in units of the coherence length $\xi(0) $ at $T=0$, the time scale is set to be GL relaxation time $t_{GL}=4\pi \lambda(0)^{2}\sigma /c^{2}$ with $\lambda(0)$ being the penetration depth at $T=0$, the temperature is in units of the superconducting critical temperature $T_c$, and the magnetic field in bulk upper critical field $H_{c2}=\Phi_{0}/2 \pi \xi(0)^{2}$  with $\Phi_{0}$ being the flux quantum. $\sigma = 1$ is the conductivity in the normal state and $\kappa$ is the GL parameter. In the present calculations, we set $\kappa = 40$ to effectively simulate a thin superconductor. With the zero electric potential gauge \cite{bennemann2008superconductivity} employed, we introduce currents through boundary conditions for magnetic field as previous simulation works \cite{PhysRevLett.109.057004,PhysRevB.86.224504,PhysRevB.103.014502}, where the unit of the current density is $J_{GL}=\sigma \hbar /2e\xi(0) t_{GL}$.

Since we are interested in the diode effect where the applied current is close to the critical current, the vortex velocity can be high, leading to significant Joule heating. The present TDGL simulations will be performed in combination with the heat-transfer equation in the following form~\cite{PhysRevB.71.184502}:
\begin{equation}
\nu \frac{\partial T}{\partial t}=\zeta \nabla^{2}T + \sigma( \frac{\partial
\mathbf{A}}{\partial t})^{2} - \eta (T - T_{0}), \label{heat}
\end{equation}%
where $T_{0}$ is the bath temperature, $\nu=0.03$ is the heat capacity of the superconducting film, and $\zeta=0.06$ the heat conductivity of the sample. The second term on the rhs of Eq. (\ref{heat}) counts for the Joule heating while the third term describes the heat exchange between the sample and its holder with an efficiency $\eta = 2 \times 10^{-4}$. Values of these phenomenological coefficients correspond to intermediate heat removal \cite{PhysRevB.71.184502} and have been proved valid for describing typical experiments \cite{lyu2021superconducting,PhysRevLett.109.057004}. All equations are solved using a semi-implicit algorithm \cite{WINIECKI2002127} with the periodic boundary condition along the direction of the external current and Neumann boundary conditions in the transverse direction.

With the assumption of an ultrathin superconducting film, we confine the TDGL simulations in two dimensions at finite temperature ($T_{0}=0.90T_c$). Without loss of the generality, we consider a strip of $L_x=200\xi (0)$ in length and $L_y=150\xi (0)$ in width. One row of MDs are aligned along the middle line of the sample as displayed in Fig. \ref{fig_sys}. Another two rows of MDs are placed symmetrically on the two sides of the central row. The distances between the nearest-neighboring rows is $h=16\xi(0)$. The magnetization of all the MDs is set to be $M$ in units of $H_{c2}$. It is found that the MD spacing in $x$ direction does not qualitatively change our main results. It is empirically adopted as $13.3\xi (0)$ in this work so that there are a total of $15$ columns of MDs.

\section{Concept and validation of the SDE}

\begin{figure}[htbp]
\centerline{
\includegraphics[scale=0.42]{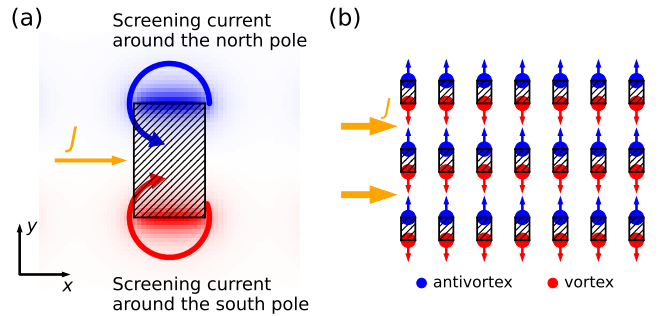}}
\caption{(a) Nucleation of VAPs driven by external currents. The background shows the distribution of the magnetic field ($z$ component) of the MD. The shaded block indicates the geometry of the MD. { The $x$ and $y$ directions indicated here  are the same as those in Fig. \ref{fig_sys}, which are also shared by all the following figures concerning the superconducting film}. (b) Illustration for the annihilation of VAPs. The Lorentz forces applied on the vortices and antivortices have the opposite direction (small colored arrows). }
\label{fig_concept}
\end{figure}

First of all, we introduce the main idea of the proposed SDE. A MD can be considered as a magnetic dipole which generates a strongly localized magnetic field. Its stray field  induces screening currents in the superconductor around the poles, as illustrated in Fig.\ref{fig_concept}(a). These currents have the opposite vorticity and can therefore be amplified in the inner region of the MD. The total superposed current can even exceed the superconducting depairing limit and generate VAPs if the magnetization of the MD is strong enough. However, we set the magnetization to be just below the  critical value so that the VAPs cannot be induced by the MDs themselves. In this case,  the nucleation of the VAPs can be controlled by external currents. With a carefully designed array of MDs, a VAP lattice can be formed with an applied current of a specific polarity [Fig.\ref{fig_concept}(b)]. These VAPs can be depinned by the Lorentz force and annihilate each other when they meet, during which the superconductivity may be destroyed due to heat dissipation. In the following, we  show the validation of the concept with detailed simulation results.

\begin{figure}[tp]
\centerline{
\includegraphics[scale=0.43]{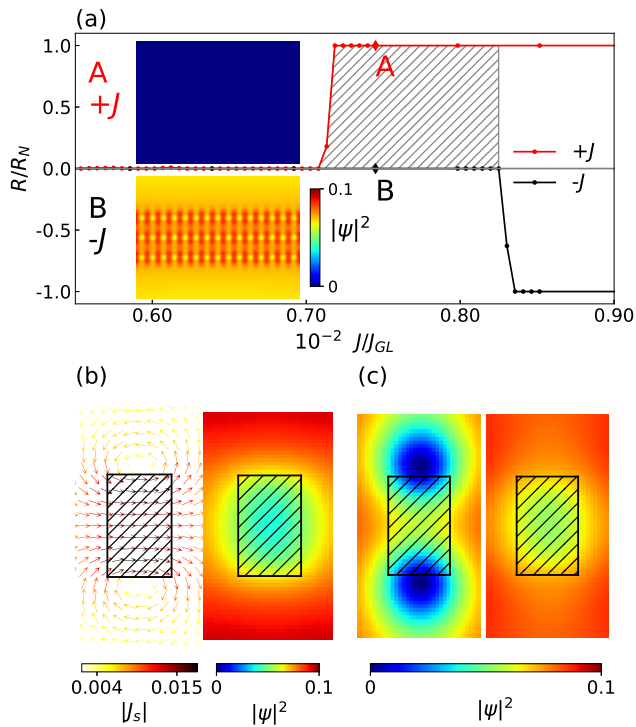}}
\caption{Simulation results at $M=0.17H_{c2}$. (a) Current-resistance characteristics. The  resistance is normalized to the resistance in the normal state $R_{N}$ and current to the GL current density $J_{GL}$. The SDE emerges in the shaded area. Insets gives the distribution of the CPD at points A and B, where the applied currents are equal but in the opposite direction. The MD array is not plotted for a clean  view of the CPD. (b) Distribution of supercurrent  (left) and CPD (right) in equilibrium, i.e., $J=0$. Direction and strength of the local supercurrent is indicated by colored arrows. The MDs are denoted by shaded blocks. (c) CPDs when the external currents are $J=+0.0069$ (left) and $-0.0069$ (right). VAPs are pinned under the poles as shown in the left panel.}
\label{fig_ref2}
\end{figure}

To begin with, we simulate the current-resistance characteristics of the present hybrid system with magnetization $M=0.17H_{c2}$ for two currents with opposite directions. As shown in Fig. \ref{fig_ref2}(a), the critical current densities $J_{c}$s where the superconductor-metal transition happens are considerably different for two currents with opposite directions $\pm J$. The sample turns to the normal state for a bias current density $J=J_{c}^{+}\simeq +0.0071$, while it remains in the superconducting state for negative currents up to $J=J_{c}^{-}\simeq -0.0083$, giving a critical current difference $\Delta J_{c}=0.0012$. That is, this system exhibits SDE with zero resistance in only one current direction. The Cooper-pair-density (CPD) distributions in the insets of Fig. \ref{fig_ref2}(a) reveal the distinct phases at opposite currents $J = \pm 0.0075$. In particular, the flux-flow state, where moving vortices yield finite voltage and partially destroy superconductivity, is absent, indicating that the present system can directly switch between superconducting and normal conducting states right at the critical currents.

To understand the underlying mechanism we further investigate details of various physical quantities in the dynamical process. We first figure out the supercurrent and the CPD distributions in the sample. Due to the presence of MDs, the Meissner current is induced around the MD, which could partially cancel the stray field. Its vorticity depends on the direction of the local field. For the present system, the distribution of the intrinsic Meissner currents around a single MD is shown in the left panel of Fig. \ref{fig_ref2}(b) in equilibrium. Owing to the dipolelike stray field of the in-plane MD, the total current in the center regime of the MD is the superposition of the roughly same directed intrinsic currents induced by the poles of opposite polarities. These currents flowing along $+x$ direction are significantly larger than anywhere else and therefore suppress local superconductivity. As demonstrated in the right panel of Fig. \ref{fig_ref2}(b),  we indeed observe a lower CPD in the center area of the MD. Note that the induced Meissner current is close to but still lower than the depairing current, so the VAPs cannot nucleate in this case.

Due to the asymmetric distribution of the MD-induced supercurrents in equilibrium, the transport property would strongly depend on the polarity of the applied current. When we apply a current $+J$ along $x$ direction, the external current enhances (suppresses) the total current in the inner (outside) region of the MDs. In contrast, an opposite current, $-J$, suppresses (enhances) the total current in the inner (outside) region. As shown in Ref. ~\cite{PhysRevLett.94.227001}, the generation of a VAP is governed by supervelocity $v \propto J_{s}/|\psi|^2$, where $J_{s}$ is the supercurrent density and $|\psi|^2$ is the CPD. Therefore, for $+J$, VAPs are expected to nucleate in the inner region with the highest supercurrent density and the smallest CPD. As shown in Fig. \ref{fig_ref2}(c), VAPs indeed form at $J=+0.0069$ (left), while they are absent at $J=-0.0069$ (right). This means that the VAPs are formed earlier for the positive currents.

\begin{figure*}[htbp]
\centerline{
\includegraphics[scale=0.37]{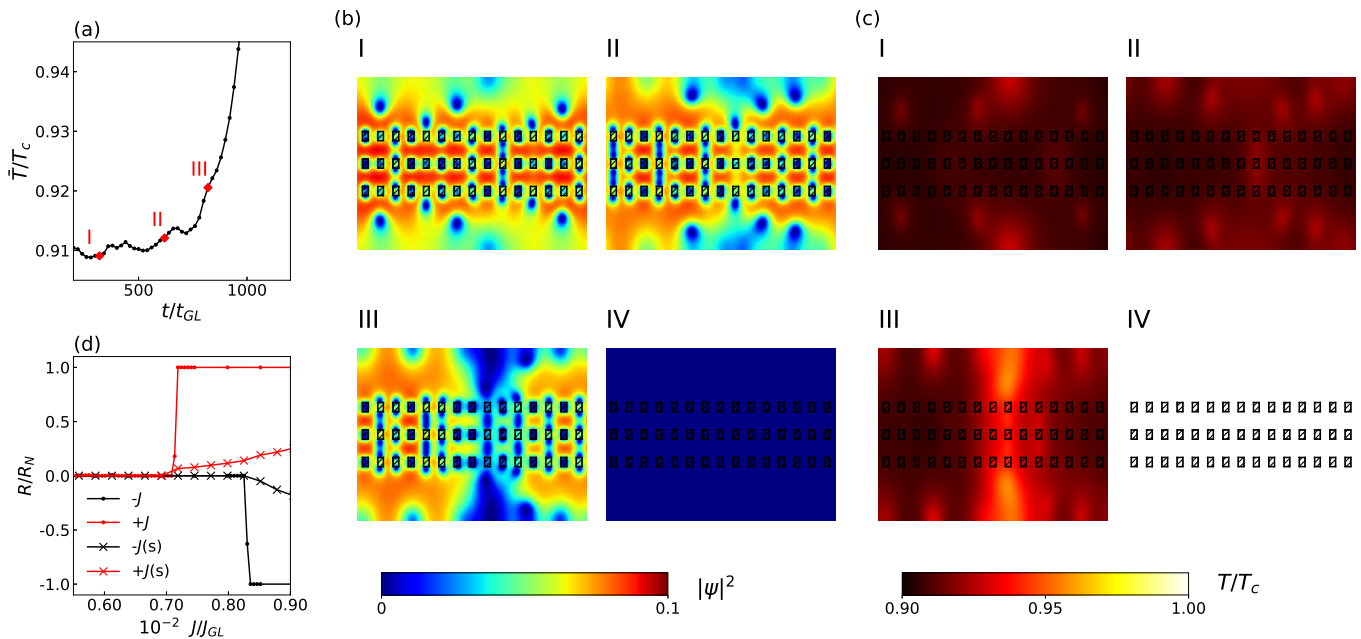}}
\caption{(a) Evolution of the mean temperature $\bar{T}$ at the critical current $J_c^+=0.0071$. The red dots marked with I, II, and III represent three typical moments during superconductor-metal transition. The black curve eventually exceeds the critical temperature, indicating that a normal state can be reached finally. (b) Snapshots of CPD at the three moments I, II and III. Image IV is the snapshot of the normal state when $t=1500t_{GL}$. We use black blocks to indicate the positions of the MDs. (c) Snapshots of the corresponding temperature distribution at the four moments. (d) I-V characteristics of systems with intermediate heat removal (dotted) and strong heat removal (marked with crosses).}
\label{fig_ref3}
\end{figure*}

The formation of VAPs is the key effect to understand the SDE. If the applied current is not strong enough, VAPs are pinned by the MD. With the increase of the external current, the Lorentz force eventually set the VAPs in motion and drive the system to dissipation state. The above picture of vortex excitations does not allow for the flux-flow state (partial superconductivity) with a steady dissipation state. Our simulations suggest that the sudden switch from the superconducting state to the normal state originates from the Joule heating caused by the motion of the VAPs.

We examine the vortex dynamics during the period when the system switches from superconducting to normal conducting. The evolution of the mean temperature is plotted in Fig. \ref{fig_ref3}(a) and the corresponding vortex images at various moments are given in (b). With a current of $J=J_c^{+} \simeq +0.0071$, the VAPs are depinned due to the Lorentz force. They are driven apart and annihilate with each other  inside the sample, or moving out of the sample.  Fig. \ref{fig_ref3}(b) gives  some snapshots for such VAP motions. The VAPs periodically nucleate and annihilate, during which the sample is gradually heated (I $\rightarrow$ II $\rightarrow$ III) while the superconductivity is suppressed (II $\rightarrow$ III). The Joule heating raises the temperature of the whole sample, which eventually exceeds the superconducting critical temperature (III $\rightarrow$ IV). Consequently the system cannot maintain superconductivity and switches to the normal state eventually (IV).

Vortex dynamics and evolution of the temperature distribution of the whole sample is visualized in Animation-S1 (right panel) within the supplemental materials~\cite{supplemental}, which would help to understand the implication between the vortex dynamics and the heat dissipation. To demonstrate the effect of Joule heating in a more direct way, we additionally conduct simulations of the same system but with a stronger heat removal $\eta=2.0 \times 10^{-3}$, which is 10 times larger. It corresponds to an extreme condition in experiments \cite{PhysRevB.71.184502} and can shed insights into  the effect of heat dissipation numerically. As shown in Fig. \ref{fig_ref3}(d), the diode effect is greatly reduced. Once  $J_c$ is reached, system enters the dissipation state rather than the normal state. A  stronger heat removal can cool down the sample more efficiently, mitigating the Joule heating caused by vortex motion and the VAP annihilation. Compared to an intermediate heat removal, a stronger heat removal suppresses the diode effect. To achieve an ideal SDE, the heat removal should not be too strong.

Next, we turn to the case of the opposite external current. A negative current reduces the supercurrent between the MD poles because they have the opposite directions. In fact, by comparing the CPD between the right panel of Fig. \ref{fig_ref2}(b) and (c), we find the superconductivity is enhanced as the low density region (colored in yellow) shrinks after applying $-J$. As a result, formation of VAPs no longer favors the regime beneath the MDs. Due to surface pinning, VAPs are expected to nucleate on sample edges first. As shown in Animation-S2 within the Supplemental Material \cite{supplemental}, at $J=J_c^{-} \simeq -0.0083$, the VAPs show near the edges first. These periodically generated VAPs bring dramatic Joule heating and drive the system to normal state with the mechanism similar to that of $+J$. Therefore a flux-flow state is absent at both $\pm J$ and an ideal SDE occurs.

We stress that the mechanism discussed above is completely different from the vortex ratchet effect, which has been studied also for  nonreciprocal transport properties \cite{jiangdong2016apl,PhysRevB.103.014502,PhysRevB.93.064508,PhysRevB.91.184502,PhysRevLett.98.117005}. Generally, in those systems, vortices are generated with an external flux applied inside the sample. The ratchet signal is mainly caused by different vortex velocities within the same phase, i.e., the flux-flow state. The corresponding ratchet voltage is thus much weaker than the present one which originates from the transition between the superconducting and normal phases.

\section{Reversed SDE for higher magnetization of MDs}

The diode effect discussed above originates from the excitation and dissipative motion of VAPs. It requires a strong magnetization to keep the local Meissner current close to the field-dependent depairing currents, so that the external current $+J$ can be used as a trigger for generating VAPs in the inner region of MDs, which results in a significantly smaller critical current in one current direction. Thus it is also interesting to see how the diode effect behaves at a strong magnetization where the MD-induced Meissner current itself exceeds the critical value for the nucleation of VAPs.

\begin{figure}[htb]
\centerline{
\includegraphics[scale=0.43]{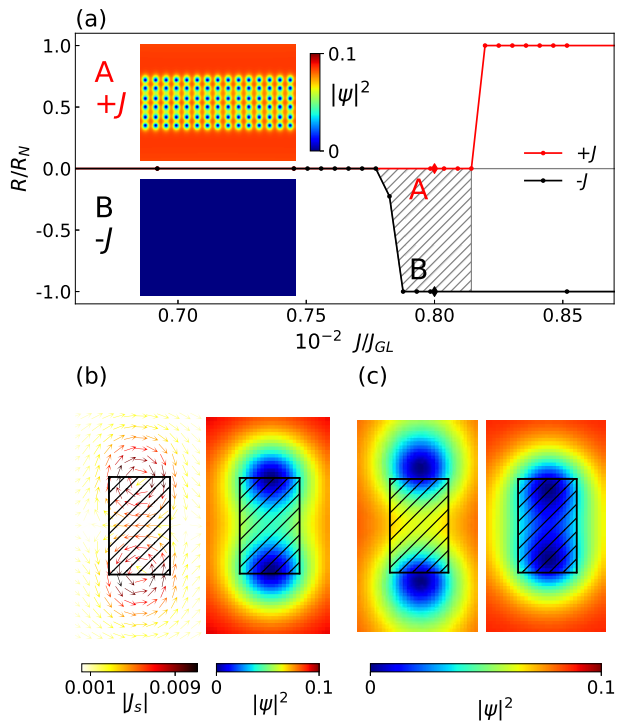}}
\caption{(a) I-R characteristics at $M=0.22H_{c2}$. The shaded region now suggests the diode effect is reversed, in contrast to that of \ref{fig_ref2}(a). The insets are visualization of CPD between two phases $J=\pm 0.0080$, which are denoted by A and B, respectively. (b) Distribution of the supercurrent and the CPD in equilibrium ($J=0$). VAPs are formed due to a strong magnetization. The shaded blocks indicate the position of MD. (c) Vortex images with currents $J=+0.0069$ (left) and $-0.0069$ (right). The balance position of VAPs depend on the polarity of the external current.}
\label{fig_ref4}
\end{figure}

We take a larger magnetization $M=0.22H_{c2}$ as an example. The I-R characteristics is presented in Fig. \ref{fig_ref4}(a). Surprisingly we find that $J_c^{-}$ becomes smaller than $J_c^{+}$, contrary to that at $M=0.17H_{c2}$ discussed above. In spite of the signal reversal, the flux-flow state is also absent. The CPD distributions given by the insets of Fig. \ref{fig_ref4}(a) confirm that the reversed diode effect also originates from transition between the superconducting phase and the normal conducting phase.

To gain insight of the reversed SDE, we first examine the distribution of supercurrent and CPD at zero current. Fig. \ref{fig_ref4}(b) shows the distributions in the vicinity of one MD. Surprisingly, vortices can be formed even without external currents. The supercurrent distribution indicates the opposite vorticity of two vortices, implying the true VAP. {   Just due to the presence of the VAPs in this case, the supercurrents around the vortices and antivortices are generated to  maintain the vorticity. As a result, the direction of the supercurrent between the poles of the MD is reversed compared to Fig. \ref{fig_ref2}(b) without VAPs at zero external currents.}

Since the MDs can generate VAPs without the help of external current, the mechanism to reach the dissipation state should be different from that discussed in the last section. In Fig. \ref{fig_ref4}(c) we plot the distribution of CPD at currents $J=\pm 0.0069$ where the zero-voltage is maintained, i.e., VAPs are pinned. These VAPs feel three forces: the pinning force of MDs, the Lorentz force due to external current, and attractive vortex-antivortex interaction. To enter the flux-flow state the latter two forces have to exceed the pinning force. The circular current of a vortex is opposite to an antivortex. For $+J$, the Lorentz force tends to drive the VAPs apart, while for $-J$ it drives the VAPs closer. As a result, when $-J$ is applied, the distance between the VAP in the right panel of Fig. \ref{fig_ref4}(c) is impressively small, yielding a stronger vortex-antivortex attraction. Therefore, with a negative external current it requires a larger pinning force to pin the VAPs. In other words, the critical current of $J_c^{-}$ must be smaller in this case, which is consistent to the I-R characteristics in Fig. \ref{fig_ref4}(a). Once the VAPs are depinned from MDs, the dissipation and Joule heating due to moving vortices emerge. Following the same mechanism described in Sec. III, the system quickly turns to normal state, resulting in the reversed diode effect. Animation S3 and S4 provided within the Supplemental Materials \cite{supplemental} visualize details of the superconducting-metal transition for $+J$ and $-J$, respectively.

Intriguingly, the SDE curves in both Figs. \ref{fig_ref2}(a) and \ref{fig_ref4}(a) are quite similar to the right part of Fig. 1c in Ref. ~\cite{ando2020observation}. It is particularly interesting that sign reversals of the SDE can be even obtained by solely changing the magnitude of the MD magnetization. This feature has not been reported previously and could be very helpful in the potential device applications.

Finally,  we note that our results for $J_c^+$ and $J_c^-$  are strongly dependent on the model parameters, especially on the position of the MDs and their magnetizations. In fact, the considerable SDE can be  found in a finite range of the  magnetization. For small magnetizations, the strongest SDE, i.e., the largest difference between $J_c^+$ and $J_c^-$, can be achieved  at $M=0.17H_{c2}$, as presented in Sec. III. On the other hand, for the large magnetizations, the strongest SDE occurs at  $M=0.22H_{c2}$ as presented in Sec. IV. The position of MDs can influence the annihilation rate of the VAPs, so it is also related to the heat conductivity $\zeta$. The SDE may be further improved and enhanced by fine-tuning parameters, as demonstrated by those reported in Refs. ~\cite{daido2021intrinsic,PhysRevB.98.144510}.

\section{Summary}

In this work, we propose a scheme with a conventional thin superconductor film and an array of MDs without a net flux applied to achieve the SDE. Due to the presence of the same number of vortices and antivortices, the mechanism for the SDE in the conventional superconductors revealed in this work is essentially different from those under a net external field. Combining the TDGL theory with the heat-transfer equation, we extensively simulate this hybrid system with finite magnetizations of the MDs. As a result, we observe the SDE without external magnetic fields. Interestingly, the SDE can be reversed by simply changing the magnitude of the MD magnetization. Moreover, the present SDE is achieved by the sudden switch between the superconducting state and normal state, excluding the flux-flow state, exhibiting a pronounced nonreciprocal voltage signal.

The topological excitations of the VAPs and their motion are responsible for the observed SDE without perpendicular net fields. The embedded MDs produce highly localized stray fields. Such dipolelike fields induce screening currents around poles which can therefore enhance each other when they meet in the center area of the MDs, and then create a strong supercurrent density between the poles. For systems with magnetization near a critical value, a positive external current ($+J$) can trigger the generation and current-driven motion of VAPs earlier than the negative one ($-J$), resulting in a polarity-dependent critical current. The motion of VAPs not only causes the dissipation voltage, but also inevitably leads to Joule heating. By constantly annihilating inside the sample and near the edges, the VAPs generate significant heat and drive the system to the normal state, resulting in an ideal SDE.

The reversed SDE can also be generated by the dynamics of VAPs in a different way. By increasing the magnetization of the MDs, VAPs are generated and pinned under the poles even in the absence of applied current. However, since the Lorentz forces applied on vortex and antivortex are antiparallel, $-J$ current brings these VAPs closer while $+J$ current drives them apart. Thus, a $-J$ current, where the attractive interaction between VAPs is stronger, results in a lower critical current. Due to Joule heating generated by the motion and annihilation of VAPs, the reversed SDE then occurs.

In short, we reveal a pronounced SDE in a conventional superconducting film and magnetic-dot hybrid structure. The size of the required MD arrays is accessible in modern nanoengineering. { The proposed structure can be conveniently realized in experiments. Thus it can be an economical candidate for practical superconducting diode among the dissipationless electronic devices.  We note that a field-free superconducting diode is also reported in  twisted multilayer graphenes~\cite{scammell2022theory,lin2021zerofield,diez2021magnetic}, showcasing the importance of the field-free diode effect in both fundamental research and potential applications.}

\section*{ACKNOWLEDGEMENTS}

This work is supported in part by the National Key Research and Development Program of China (No.2017YFA0303002) (Q.H.C. and J.J.) and (2018YFA0209002) (Y.L.W.); and the National Natural Science Foundation of China (Nos. 11834005, 11674285, 61771235, and 61727805). Z.L.X. acknowledges support by the U. S. National Science Foundation under Grant No. DMR-1901843. F.M.P. and M.V.M. acknowledge support by the Research Foundation - Flanders (FWO).

%

\end{document}